\input harvmac
\newcount\figno
\figno=0
\def\fig#1#2#3{
\par\begingroup\parindent=0pt\leftskip=1cm\rightskip=1cm\parindent=0pt
\baselineskip=11pt
\global\advance\figno by 1
\midinsert
\epsfxsize=#3
\centerline{\epsfbox{#2}}
\vskip 12pt
{\bf Fig. \the\figno:} #1\par
\endinsert\endgroup\par
}
\def\figlabel#1{\xdef#1{\the\figno}}
\def\encadremath#1{\vbox{\hrule\hbox{\vrule\kern8pt\vbox{\kern8pt
\hbox{$\displaystyle #1$}\kern8pt}
\kern8pt\vrule}\hrule}}

\overfullrule=0pt

%macros
%
\def\tilde{\widetilde}
\def\bar{\overline}
\def\Z{{\bf Z}}
\def\T{{\bf T}}
\def\S{{\bf S}}
\def\R{{\bf R}}

\font\zfont = cmss10 %scaled \magstep1
\font\litfont = cmr6

\def\bigone{\hbox{1\kern -.23em {\rm l}}}
\def\ZZ{\hbox{\zfont Z\kern-.4emZ}}
\def\half{{\litfont {1 \over 2}}}

\Title{hep-th/9607163, IASSNS-HEP-96-78, RU-96-55}
{\vbox{\centerline{GAUGE DYNAMICS}
\bigskip
\centerline{AND COMPACTIFICATION}
\bigskip
\centerline{ TO THREE DIMENSIONS }}}
\smallskip
\centerline{\it Nathan Seiberg\foot{Supported in part by
DOE DE-FG02-96ER40559} }
\smallskip
\centerline{\it Department of Physics and Astronomy}
\centerline{\it Rutgers University }
\centerline{\it Piscataway, NJ 08855-0849}
\bigskip
\centerline{and}
\bigskip
\centerline{\it Edward Witten\foot{Supported in part by
NSF PHY95-13835} }
\smallskip
\centerline{\it School of Natural Sciences, Institute for Advanced Study }
\centerline{\it Olden Lane, Princeton NJ 08540}
\bigskip

\medskip

\noindent
We study four dimensional $N=2$ supersymmetric gauge theories on $R^3
\times S^1$ with a circle of radius $R$.  They interpolate between 
four dimensional gauge theories ($R=\infty$) and $N=4$ supersymmetric
gauge theories in three dimensions ($R=0$).  The vacuum structure
can be determined quite precisely as a function of $R$, agreeing
with three and four-dimensional results in the two limits.

\Date{June, 1996}

\newsec{Introduction}

\nref\sw{N. Seiberg and E. Witten, ``Electric-Magnetic Duality, Monopole
Condensation, And Confinement In $N=2$ Supersymmetric Yang-Mills Theory,''
Nucl. Phys. {\bf B426} (1994) 19.}%
\nref\secondsw{N. Seiberg and E. Witten, ``Monopoles, Duality, And Chiral
Symmetry Breaking In $N=2$ Supersymmetric QCD,'' Nucl. Phys. {\bf B431} 
(1994)  484.}%

In \refs{\sw, \secondsw}, the dynamics of the Coulomb branch of $N=2$
super Yang-Mills theory was analyzed using general constraints of
supersymmetry and low energy effective field theory -- extended,
crucially, by allowing for the possibility of duality transformations.
The purpose of the present paper is to study the same theory
compactified or reduced to three dimensions.

Compactification
to three dimensions means that one formulates the quantum theory on
$\R^3\times \S^1_R$,
where  $\S^1_R$ is a circle of circumference $2\pi R$.  For $R\to \infty$
one should recover the four-dimensional solution of \refs{\sw, \secondsw}.

Dimensional reduction means instead that at the classical level, one
takes the fields to be independent of the fourth dimension, and then
one quantizes the resulting three-dimensional theory.  Intuitively,
one would expect that this three-dimensional theory should be
equivalent to the small $R$ limit of compactification.  After all, the
energetic cost of excitations that carry non-zero momentum along
$\S^1_R$ diverges as $R\to 0$.

\nref\sen{A. Sen, ``$F$-Theory And Orientifolds,'' hep-th/9605150.}%
\nref\bds{T. Banks, M. Douglas, and N. Seiberg, 
``Probing $F$-Theory With Branes,'' hep-th/9605199.}%
\nref\seiberg{N. Seiberg, ``IR Dynamics On Branes And Space-Time Geometry,''
hep-th/9606017.   }%
In section two of this paper, the Coulomb branch of the three-dimensional
theory will be analyzed, for gauge groups $SU(2)$ and  $U(1)$.
In fact, drawing upon ideas of \refs{\sen,\bds},
results on this subject have been inferred recently {}from
string theory \seiberg.
Here we will show what can be learned about the problem
using some simple arguments of field theory, and in particular
we recover many of the results of \seiberg.
In section three, we analyze the four-dimensional quantum theory on
$\R^3\times \S^1_R$ using some simple field theory arguments, among
other things 
verifying that the large $R$ limit gives back the four-dimensional
theory while the small $R$ limit gives the three-dimensional theory.
In section four we recover  and explain results of section three {}from 
the standpoint of string theory.

\newsec{The Three-Dimensional Theory}

\subsec{ The Problem}

We will here be discussing three-dimensional supersymmetric gauge
theories which have $N=4$ supersymmetry in the three-dimensional sense
(corresponding to $N=2$ in four dimensions).  They can be constructed
by dimensional reduction of six-dimensional $N=1$ super Yang-Mills
theory to three dimensions.  This is a convenient starting point in
understanding the field content and symmetries of the models.  First
we consider the pure gauge theories, without matter hypermultiplets.

In six dimensions, the fields are the gauge field $A$ and
Weyl fermions $\psi$ in the adjoint representation of the gauge
group $G$.  There is an $SU(2)_R$ symmetry that acts only
on the fermions; the fermions and
supercharges transform as  doublets of $SU(2)_R$.

Upon dimensional reduction to three dimensions -- that is, taking
the fields to be independent of three coordinates $x^{4,5,6}$ --  
one obtains a theory with the following additional structures.
The last three components of $A$ become in three dimensions
scalar fields $\phi_i$, 
$i=1,2,3$, in the adjoint representation.  These scalars
transform in the vector representation under the group
 of rotations of the $x^{4,5,6}$; we will call the double cover
of this group $SU(2)_N$.  Note that in reduction
to four dimensions, only two such scalars appear, and instead
of $SU(2)_N$, one gets only a $U(1)$ symmetry of rotations of
the $x^{5,6}$ plane.  This symmetry is often called $U(1)_R$, and
has an anomaly involving four-dimensional instantons.  In three
dimensions, because the group $SU(2)_N$ is simple, there is no
possibility of such an anomaly.  Finally,  three
dimensional Euclidean space $\R^3$ has a group  of rotations whose
double cover we will call $SU(2)_E$.

Under $SU(2)_R\times SU(2)_N\times SU(2)_E$, the fermions transform
as $({\bf 2},{\bf 2},{\bf 2})$, as do the supercharges (so that
$SU(2)_N$ is a group of $R$ symmetries just like $SU(2)_R$),
while the scalars transform as $({\bf 1},{\bf 3},{\bf 1})$.

Now to formulate the problem of the Coulomb branch, the starting
point is the potential energy for the scalars.    This arises
by dimensional reduction {}from the $F^2$ kinetic energy of gauge
fields in six dimensions, and is
\eqn\potential{V={1\over 4e^2}\sum_{i<j}\Tr [\phi_i,\phi_j]^2}
where $e$ is the gauge coupling.
For the classical energy to vanish,  it is necessary
and sufficient that the $\phi_i$ should commute.  One can consequently
take them to lie in a maximal commuting subalgebra of the Lie
algebra of $G$.  If $G$ has rank $ r$, the space of zeroes
of $V$, up to gauge transformation, has real dimension $3r$.
A generic set of commuting $\phi_i$ breaks $G$ to an Abelian
subgroup $U(1)^r$.  In addition to the $\phi_i$, there are
then $r$ massless photons.  Since a photon is dual to a scalar
in three space-time dimensions, there are in all $4r$  massless
scalars -- $3r$ components of $\phi_i$ and $r$ duals of the photons.

Are these $4r$ scalars really massless in the quantum theory?
The $N=4$ supersymmetry makes it impossible to generate a superpotential,
so there are only two rather special ways to have masses.
One possibility is to include a three-dimensional Chern-Simons
interaction, with a quantized integer-valued coupling $k$.  For
non-zero $k$, the modes described above do indeed get masses,
and the problem we will pose in this paper of studying the Coulomb
branch does not arise.  (There is an interesting question of whether
the theory with $k\not= 0$ has a supersymmetric vacuum; at least for
large $k$, the answer can be seen to be ``yes'' by using perturbation
theory in $1/k$.)   If the gauge group $G$ has $U(1)$ factors,
it is possible to include Fayet-Iliopoulos $D$-terms (transforming
as $({\bf 3},{\bf 1},{\bf 1})$ under $SU(2)_R\times SU(2)_N\times 
SU(2)_E$), again giving mass to some modes.  In this paper, we will
mainly consider the case that $G$ is semi-simple, so that $D$-terms
are impossible; but even when we consider $G=U(1)$, we will focus
on the case that the $D$-terms are absent.  

With these restrictions, then, the $4r$ scalars are really massless
and parametrize a family of vacuum states.  (This is also true later
when we include hypermultiplets.)  Moreover, by considering 
the region of large $\phi_i$, we know that for a generic vacuum in this 
family, the physics is free in the infrared and can be described by a 
conventional low energy effective field theory.
The most general low energy effective action for $4r$ massless scalars in
three dimensional $N=4$ supersymmetry is a sigma model with
a target space that is a hyper-Kahler manifold of quaternionic
dimension $r$.  Thus, the moduli space ${\cal M}$ of vacua is
to be understood as such a hyper-Kahler manifold.  

\nref\ah{M. F. Atiyah and N. Hitchin, {\it The Geometry And Dynamics
Of Magnetic Monopoles} (Princeton University Press, 1988).}%
In this paper, we will only consider in detail the cases $G=SU(2)$ 
and $G=U(1)$, for which $r=1$, and ${\cal M}$ is simply a 
hyper-Kahler manifold of real dimension four.  Moreover,
this manifold has a non-trivial action of $SU(2)_N$,
which highly constrains the problem; the hyper-Kahler manifolds
we need are (with one easy exception, the reason for which will emerge)
 to be found in the classification in \ah\ of certain four-dimensional
hyper-Kahler manifolds with $SO(3)$ symmetry.  

\nref\ewitten{E. Witten, ``An $SU(2)$ Anomaly,'' Phys. Lett. {\bf B117} 
(1982) 324.}%
\nref\aw{L. Alvarez-Gaum\'e and E. Witten, ``Gravitational Anomalies,''
Nucl. Phys. {\bf B234} (1983) 269.}%
So far we have discussed the pure gauge theories.  It is also
possible to include matter hypermultiplets.  For $G=SU(2)$, we will
consider in some detail the case of matter hypermultiplets in the
doublet or
two-dimensional representation of $G$.  The basic such 
object is a multiplet
that contains four real scalars that transform as $({\bf 2},{\bf 1},
{\bf 1},{\bf 2})$
under $SU(2)_R\times SU(2)_N\times SU(2)_E\times G$,
along with fermions transforming as $({\bf 1},{\bf 2},{\bf 2},{\bf 2})$.
  For somewhat quirky reasons, such a
multiplet is sometimes called a half-hypermultiplet.  In \sw, the
$G=SU(2)$ theory was studied (in four dimensions) with any number $N_f$
of doublet hypermultiplets, or in other words $2N_f$ half-hypermultiplets.
With this notation, it appears that we should allow for the case in which
$N_f$ is a half-integer rather than an integer, but at this
point some subtleties involving global anomalies intervene.
In four dimensions, given the fermion content of the half-hypermultiplet,
the theories with half-integral $N_f$ are simply inconsistent
because of a $\Z_2$ global anomaly \ewitten. 
In three dimensions, the situation is somewhat different.
The theories with half-integral $N_f$ exist, but for those
theories the Chern-Simons coupling $k$ cannot vanish, and the
Coulomb branch that we will be studying in this paper does not exist.
In fact, because of a global anomaly (see p. 309 of \aw), 
$k$ is congruent to $N_f$
modulo ${\bf Z}$, 
and can vanish only if $N_f$ is integral.\foot{In terms
immediately relevant to this paper, the global anomaly pointed
out in \aw\ would show up as follows.  If $N_f$ is half-integral,
then the number of fermion zero modes in a monopole field would
be odd. This appears to lead to a contradiction as amplitudes
in a monopole field would change sign under a $2\pi$ rotation.
The resolution of the paradox is not that the theory does not exist,
but that when $N_f$ is odd, $k$ is half-integral and in particular
non-zero; as non-zero $k$ gives the photon a mass, finite action
monopoles do not exist.}
So we will only consider integer $N_f$ in this paper.

For the other case $G=U(1)$, we will consider the behavior with
an arbitrary number $M$ of hypermultiplets of charge one.

Until further notice, all of our hypermultiplets will have
zero bare mass.  After understanding the case of zero bare mass,
we will make brief remarks on the role of the bare masses.

\subsec{Behavior At Infinity}

The starting point of the analysis is to understand what
happens in the semi-classical region of large $|\phi|$. 

For the potential energy $V$ to vanish means that the $\phi_i$ commute 
and so can be simultaneously diagonalized by a gauge transformation.  
This means for $SU(2)$ that one can take
\eqn\simdi{\phi_i=\left(\matrix{ a_i & 0 \cr 0 & -a_i \cr}\right)}
for some $a_i$.  The $a_i$ are defined up to a Weyl transformation,
which exchanges the two eigenvalues of the $\phi_i$, and so acts
as $a_i \to -a_i$.  The space of zeroes of $V$ is thus
a copy of $\R^3/\Z_2$.  For a complete description of the moduli
space of vacua, one must also include an extra circle, parametrizing
a fourth scalar $\sigma$ which is dual to the photon.  The Weyl
group (which acts by charge conjugation) multiplies also the
fourth scalar by $-1$. So the space of vacua at the classical
level is $(\R^3\times \S^1)/\Z_2$, where the $\Z_2$ multiplies
all four coordinates by $-1$.  The classical metric on the moduli
space is a flat metric 
\eqn\flatmet{ds^2={1\over e^2} \sum_id\phi_i^2 + e^2 d\sigma^2.}
The factor of $1/e^2$ for the $\phi_i$ reflects the fact
that (like the whole classical Lagrangian) the $\phi$ kinetic
energy is of order $1/e^2$.  The photon kinetic energy is likewise
of order $1/e^2$, but after duality this turns into $e^{+2}$ for $\sigma$.
(Some constants in \flatmet, omitted in this section for simplicity,
are worked out in detail in section three.)

For $G=U(1)$, there is no Weyl group and the classical moduli
space is simply $\R^3\times U(1)$.  For simplicity and to treat
the two cases in parallel, we will postpone dividing by the Weyl
group until the end of the discussion, and formulate the following
as if classically one is on $\R^3\times \S^1$.  The region at
infinity in $\R^3$ is homotopic to a  two-sphere.  Thus,
topologically we have at infinity a product $\S^2\times \S^1$
at the classical level.  As one goes to infinity, the $\S^2$
grows (radius proportional to $|\phi|$) but the $\S^1$ has a fixed
circumference of order $e$.  The $\S^2$ is visible classically,
but the $\S^1$, which appears via duality, is a more subtle
part of the quantum story.  The possibility exists that in the quantum
theory,
instead of a product $\S^2\times \S^1$ at infinity, one has
an $\S^1$ fiber bundle over $\S^2$.  In fact, to describe such
a fiber bundle, as noted in \seiberg, the classical metric should
be changed to something like
\eqn\qmet{ds_Q^2={1\over e^2}\sum_id\phi_i^2+e^2(d\sigma-sB_i(\phi)
d\phi^i)^2,}
where here $B$ is the Dirac monopole $U(1)$ gauge field over $\S^2$,
and {\it a priori} $s$ is any integer.  Because \qmet\
differs {}from the classical metric only in terms of order $e^2$,
quantum loop corrections can be responsible for changing \flatmet\
to \qmet\ and so for generating $s\not= 0$.  In fact, if $A$ is the
undualized $U(1)$ gauge field, then the integer $s$ would show
up prior to duality in an interaction $s \epsilon^{\lambda\mu\nu}
A_\lambda\epsilon_{ijk}
\hat \phi^i\partial_\mu\hat \phi^j\partial_\nu\hat \phi^k$, where $\hat
\phi^i=\phi^i/(\phi\cdot\phi)^\half$;
because it multiplies no power of $e$, this interaction
 could arise as a one-loop effect.  

The integer $s$ could thus, as was proposed in \seiberg, be computed
{}from a one-loop diagram.  We will instead compute it mainly by counting
fermion zero modes in a monopole field.

\bigskip\noindent
{\it Non-Trivial $\S^1$ Bundles Over $\S^2$}

As background, and to help in interpreting the results, let
us recall the detailed description of non-trivial $\S^1$ bundles
over $\S^2$.  An $\S^1$ bundle over any base $B$ (with oriented
fibers) is classified 
topologically
by the  Euler class of the bundle, 
which takes values in $H^2(B,\Z)$; as $H^2(\S^2,
\Z)\cong \Z$, the possible bundles over $\S^2$  are labeled
by an integer $s$, which was introduced in \qmet.  For $B=\S^2$,
the possible non-trivial bundles may be described in the following
standard fashion.

The basic example is simply the three-sphere, regarded as a fiber
bundle over $\S^2$.  Let $u_\alpha$, $\alpha=1,2$ be two complex
numbers with 
\eqn\heffo{|u_1|^2+|u_2|^2=1.}  
The possible $u_\alpha$
parametrize a copy of $\S^3$.  If we set
\eqn\effo{\vec n=\bar u \vec\sigma u,}
with $\vec\sigma$ the usual Pauli $\sigma$ matrices, then in a standard
fashion one can show by consequence of \heffo\ that 
$\vec n^2=1$.  Thus the map {}from $u$ to $\vec n$ is a map {}from $\S^3$
to $\S^2$.  All $\vec n$'s arise, and for given $\vec n$, $u$ is unique
up to a $U(1)$ transformation 
\eqn\opol{u_\alpha\to e^{i\theta}u_\alpha,~~
0\leq \theta\leq 2\pi.}  Thus the space of $u$'s for given $\vec n$
is a copy of $U(1)=\S^1$; the map {}from $\S^3$ to $\S^2$ exhibits
$\S^3$ as a  fiber bundle over $\S^2$ with fiber $\S^1$.

To introduce an arbitrary integer $s$, we begin now with $\S^3\times
\S^1$, labeling the $\S^1$ by an angle $\psi$ ($0\leq \psi\leq
2\pi$), and divide by a $U(1)$ group that acts by
\eqn\rorty{u_\alpha\to e^{i\theta} u_\alpha,~~ \psi\to \psi+s\theta.}
Let $L_s$  be the quotient $(\S^3\times \S^1)/U(1)$ with the given
$U(1)$ action.  Then $L_s$  maps to $\S^2$ by forgetting $\psi$; as we 
have noted above, the quotient of $u$-space by $u\to e^{i\theta}u$ is 
$\S^2$.  The fiber of the map to $\S^2$ is a circle,
so $L_s$ is a circle bundle over $\S^2$, for any $s$.

Let us next work out the topology of $L_s$.  We note that $L_0$
is the trivial bundle $\S^2\times \S^1$; in this case,
the $U(1)$ in \rorty\ does not act on the second factor in $\S^3\times
\S^1$, and dividing by it projects the first factor to $\S^2$.  In
general, $L_{-s}$ is mapped to $L_s$ by $\psi\to -\psi$, so they
have the same topology.  Finally, for any $s>0$, $L_s$ is isomorphic
to the ``lens space'' $\S^3/\Z_s$ obtained by dividing $\S^3$
by $u_\alpha\to e^{2\pi i k/s}u_\alpha$, $k=0,1,\dots,s-1$.  One sees
this by using the $\theta$ in \rorty\ to ``gauge away'' $\psi$,
leaving a residual $\Z_s$ gauge symmetry that acts on $u$.

The lens space $L_s$ has a manifest $SU(2)\times U(1)$ symmetry,
where the $SU(2)$ acts in the standard fashion on the $u_\alpha$
and the $U(1)$ acts by $\psi\to \psi+{\rm constant}$.  Any circle
bundle over $\S^2$ with $SU(2)\times U(1)$ symmetry will be equivalent
to $L_s$ with some value of $s$; we want a 
practical way to determine $s$.  Suppose one is sitting at some
point on $\S^2$, say $\vec n=(0,0,1)$.  In a standard basis of the
Pauli matrices, this corresponds to $u_\alpha=(1,0)$.  The point
$\vec n=(0,0,1)$ is invariant under a $U(1)$ subgroup of $SU(2)$,
consisting of rotations about the third axis; on the $u_\alpha$ this
acts by 
\eqn\kikko{
J={i\over 2}\left(u_1{\partial\over\partial u_1} - u_2{\partial\over
\partial u_2}\right).}
The $1/2$ is present because the $u_\alpha$ are in the spin one-half
representation of $SU(2)$, and is consistent with the fact that
$e^{2\pi J}=1$ in acting on $\vec n$.
Sitting at the point $u=(1,0)$, that transformation is equivalent 
(modulo a ``gauge transformation'' \rorty) to that 
generated by 
\eqn\jikko{\tilde J=-{s\over 2}{\partial\over\partial\psi}.}  
So we get our criterion
for determining the value of $s$: {\it a rotation around a given
point  $P\in \S^2$ acts with charge $-s/2$ on the $\S^1$ fiber over $P$}.
In particular, such a  rotation  shifts $\psi$ by
$\pi s$, so that $SU(2)$ acts faithfully on $L_s$ if $s$ is odd,
but $SU(2)/\Z_2=SO(3)$ acts if $s$ is even.

Since, in the case of gauge group $G=SU(2)$, we are interested in dividing
by the Weyl group, we should also discuss $\S^1$ bundles over
${\bf RP}^2=\S^2/\Z_2$.  The transformation $ \vec n \to -\vec n$
corresponds in terms of $u_\alpha$ to 
\eqn\jury{\alpha:
(u_1,u_2)\to (\bar u_2,-\bar u_1).}  In the quantum field theories
we want to study, the Weyl group also acts
on $\psi$ (the dual of the photon) by $\alpha(\psi)= -\psi$ (and this
is in any case needed for consistency with the ``gauge invariance''
\rorty), so the
circle bundles $M_s$ over ${\bf RP}^2$ that we want are obtained simply
by dividing $L_s$ by a $\Z_2$ that acts as \jury\ on $u$ and multiplies
$\psi$ by $-1$.  
We recall that in turn $L_s=\S^3/\Z_s$, where $\Z_s$ is generated
by $\beta:u_\alpha\to e^{2\pi i/s}u_\alpha$.  
So $M_s$ is the quotient of $\S^3$ by the group
generated by $\alpha$ and $\beta$.  There is no
loss of generality in assuming  that $s$ is even, say $s=2k$, since
if $s$ is odd, by replacing the group generators $\alpha$ and $\beta$
by $\alpha$ and $\alpha\beta$, one can reduce to the even $s$ case (the
point being that if $\beta$ is of odd order, then $\alpha\beta$ is of
even order).
The group generated by $\alpha$ and $\beta$ is then a dihedral
group  $\Gamma_k$  characterized by the relations
\eqn\therel{\eqalign{\alpha^2=\beta^k=-1 \cr
                     \alpha\beta =\beta^{-1}\alpha,\cr}}
where in the first relation $-1$ (which in our realization of the
group acts by $u_\alpha\to -u_\alpha$) is understood as a central
element of $\Gamma_k$.  In the correspondence between finite
subgroups of $SU(2)$ and the $A-D-E$ series of Lie groups, the
group $\Gamma_k$ corresponds to $D_{k+2}$, that is, to $SO(2k+4)$.  

\subsec{Behavior In A Monopole Field}

\nref\poly{A. M. Polyakov, ``Quark Confinement And The Topology Of Gauge 
Groups,''
Nucl. Phys. {\bf B120} (1977) 429.}
One of the key aspects of $2+1$ dimensional gauge theories is that,
as first explained by Polyakov twenty years ago \poly, magnetic
monopoles in unbroken $U(1)$ subgroups of the gauge group can appear
as instantons.

\def\C{{\bf C}}
\nref\ahw{I. Affleck, J. A. Harvey, and E. Witten, ``Instantons And
(Super)Symmetry Breaking In $2+1$ Dimensions,'' Nucl. Phys. 
{\bf B206} (1982) 413.}%
\nref\oldseiberg{N. Seiberg, ``Supersymmetry And Non-Perturbative Beta
Functions,'' Phys. Lett. {\bf 206B} (1988) 75.}%
\nref\beckers{K. Becker, M. Becker, and A. Strominger, 
``Fivebranes, Membranes, And Nonperturbative String Theory,'' Nucl.Phys. 
{\bf B456} (1995) 130. }%
The contribution of such an instanton is obviously proportional
to $e^{-I}$, where $I$ is the action of the instanton.  A more
subtle fact is that \poly\ if $\sigma$ is the scalar dual to the
$U(1)$ gauge field, then the instanton contribution also has a factor
of $e^{-i\sigma}$, incorporating in the dual description the long
range fields of the instanton.  Beyond these general factors
of $e^{-(I+i\sigma)}$, there may be additional factors coming,
for instance, {}from fermion zero modes. For example \ahw, in $N=2$ super 
Yang-Mills theory, with the instanton being a solution of the 
Bogomol'nyi-Prasad-Sommerfeld (BPS) monopole equation,
the instanton is invariant under half of the four supercharges; the others
generate two   fermion zero modes.  The field $I+i\sigma$ is the bosonic
part of a chiral superfield.  The effect of the fermion zero modes
is that the function $e^{-(I+i\sigma)}$ must be integrated over
chiral superspace, and is a superpotential rather than an ordinary
potential.  

In the present context of $N=4$ super Yang-Mills theory,  there are
eight supercharges, of which half annihilate a
supersymmetric instanton.
As in \refs{\oldseiberg,\beckers}, a supersymmetric solution in 
such a context will (if additional fermion zero modes are absent or
can be absorbed) generate a correction to the metric on moduli space,
rather than a superpotential.  We first consider the minimal
$N=4$ theory, without hypermultiplets, in which the fermion
zero modes are generated entirely by the unbroken supersymmetries.

\def\2{{\bf 2}}
\def\1{{\bf 1}}
As usual in instanton physics, it is essential to analyze the
symmetries of the instanton amplitude.  We recall that the $N=4$
gauge theory in three dimensions has a symmetry group $SU(2)_R
\times SU(2)_N\times SU(2)_E$, with the supercharges transforming
as $(\2,\2,\2)$.  The BPS monopole is invariant under the rotation
group $SU(2)_E$ (mixed with a gauge transformation) and under
$SU(2)_R$ (which only acts on fermions).  However, the choice of a vacuum
expectation value of the $\phi_i$ breaks $SU(2)_N$ to a subgroup
$U(1)_N$ even before one considers monopoles; the BPS monopole
is constructed using only a single real scalar in the adjoint,
which can be chosen to be the field with an expectation value at
infinity, and so the BPS monopole is invariant under $U(1)_N$.  

Under the unbroken group $SU(2)_R\times SU(2)_E\times U(1)_N$,
the supercharges transform as $(\2,\2)^{1/2}\oplus (\2,\2)^{-1/2}$,
where the superscript is the $U(1)_N$ charge, which takes
half integral values on the supercharges because they transform
as spin one-half under $SU(2)_N$.
The BPS monopole is invariant under half of the supercharges
in an $SU(2)_R\times SU(2)_E\times U(1)_N$-invariant 
fashion, so the unbroken
supersymmetries must be, if we pick the sign of the $U(1)_N$ generator
appropriately, the piece transforming as $(\2,\2)^{-1/2}$.  The
fermion zero modes therefore have the quantum numbers $(\2,\2)^{1/2}$.
The instanton amplitude is  schematically
\eqn\inamp{\psi\psi\psi\psi~ e^{-(I+i\sigma)},}
where the $\psi$'s are fermions of $U(1)_N=1/2$.
Note that if we consider antimonopoles instead of monopoles, 
the zero modes transform as $(\2,\2)^{-1/2}$, and \inamp\ is replaced by
\eqn\timamp{\tilde\psi\tilde\psi\tilde\psi\tilde\psi~ e^{-(I-i\sigma)},}
with $\tilde\psi$ being fermions of $U(1)_N=-1/2$.

The $\psi\psi\psi\psi$ vertex carries $U(1)_N$ charge $4\cdot (1/2)=2$.
One might be tempted to conclude that there is an anomaly in $U(1)_N$
conservation in a monopole field, 
but this is impossible as $U(1)_N$ is a subgroup
of the simple group $SU(2)_N$.  Rather, we must assign a transformation
law to $\sigma$ so that the instanton amplitude is invariant.
Clearly, this means that the $U(1)_N$ generator must act on $\sigma$ as
$+2\partial/\partial\sigma$, meaning
that in the notation \jikko\ (including the factor of $1/2$ present
there), $s=-4$ for the
pure $N=4$ gauge theory.   The moduli space of the pure $N=4$
theory therefore does not look at infinity like $\S^2\times \S^1$
but like the lens space $L_{-4}$ described in the last subsection.

\nref\callias{C. Callias, ``Index Theorems On Open Spaces,'' Commun.
Math. Phys. {\bf 62} (1978) 213.}
Now, let us determine the value of $s$ if one includes hypermultiplets
in the  two-dimensional representation of $SU(2)$.  A doublet
half-hypermultiplet
in a  monopole field has a single fermion zero mode (for the relevant index
theorem see \callias), with the opposite
sign of $U(1)_N$ {}from that of the  vector multiplet zero modes.
So with $N_f$ hypermultiplets ($2N_f$ half-hypermultiplets), there
are $2N_f$ zero modes, giving\foot{It is curious that in four-dimensional
$N=2$ super Yang-Mills theory, the analogous counting of zero
modes in an instanton field gives a factor of $-8+2N_f$, instead
of $-4+2N_f$.    The difference
arises because the half-hypermultiplet has the same number
of  fermion zero
modes in a three-dimensional monopole or four-dimensional instanton, but
 the vector multiplet has twice 
as many zero modes in the four-dimensional case
 -- four generated by ordinary supersymmetries that have an
analog in the three-dimensional problem,
 and four more by superconformal symmetries that do not.}
\eqn\jingo{s=-4+2N_f.}

For future use, we can also now work out the value of $s$
for a $U(1)$ theory with hypermultiplets.  There are no monopoles in 
the pure $U(1)$ gauge theory, but by thinking of $s$ as the coefficient
of a one-loop amplitude, and the fields of the $U(1)$ theory
as a subset of the fields of an $SU(2)$ theory, one can infer
the result for $U(1)$ {}from that for $SU(2)$.
The $U(1)$ theory without hypermultiplets is free, so the 
vector multiplet contributes nothing.  
The hypermultiplet contribution in the $SU(2)$ theory with doublet
hypermultiplets can be inferred {}from a one-loop diagram with 
the hypermultiplet running around the loop and external fields
being vector multiplets.  If we simply restrict the external
fields to be in a $U(1)$ subalgebra, then the $SU(2)$ diagram
with the internal fields being a doublet half-hypermultiplet
turns into the $U(1)$ diagram with the internal fields being
a hypermultiplet of charge one.
(In particular,
 if we embed $U(1)$ in $SU(2)$ so that the doublet of $SU(2)$ 
has $U(1)$ charges $\pm 1$, then a half-hypermultiplet of $SU(2)$ reduces
to  an ordinary charge one hypermultiplet of $U(1)$.)  
The value of $s$ 
for a  $U(1)$ theory with $M$ hypermultiplets of charge 1 is thus 
obtained by replacing $4$ by 0 and $2N_f$ by $M$ in \jingo:
\eqn\dokn{s=M.}

Going back to the $SU(2)$ theory, we see {}from \jingo\
 that $s$ is always even.   This means (as noted following \jikko)
that it is not $SU(2)_N$ but $SU(2)_N/\Z_2$, which we will call
$SO(3)_N$, that acts faithfully on the moduli space ${\cal M}$ of vacua.
Furthermore, $s\not=0$ except for $N_f=2$.  When $s\not=0$, $SO(3)_N$
acts non-trivially on the scalar $\sigma$ that is dual to the photon.
This means that the generic $SO(3)_N$ orbit is three-dimensional.
Also, because $SU(2)_N$ is a group of $R$ symmetries, the three
complex structures of the hyper-Kahler manifold ${\cal M}$ are rotated
by the $SO(3)_N$ action.
In \ah, four-dimensional hyper-Kahler manifolds with an $SO(3)$ action
that rotates the complex structures and has generic three-dimensional
orbits were classified.
{}From what has just been said, all of our metrics will appear
on their list except for $N_f=2$.  

\subsec{The Metric On Moduli Space}

Before comparing to results of \ah, and to expectations {}from
string theory, let us ask what sort of metrics
we expect on the moduli space ${\cal M}$, for various  $N_f$.
First we consider the case of gauge group $SU(2)$.
The starting point is the classical answer, the flat metric on
$(\R^3\times \S^1)/\Z_2$.  There is then a one loop correction
to the structure at infinity, for $N_f\not= 2$. The effect of this
correction is that ``infinity'' for $N_f\not= 2$ looks not like $(\S^2
\times \S^1)/\Z_2$ but like $L_s/\Z_2$, with $s=2N_f-4$.  

Perturbation corrections to the metric on ${\cal M}$ are entirely
determined by the one-loop correction plus the non-linear terms
in the Einstein equations.  (This is analogous to the fact that in four
dimensions, perturbative corrections beyond one loop are forbidden
by holomorphy.)  This may be proved as follows.
A ``new'' $k$-loop correction to the metric would
be a self-dual
solution of the linearized Einstein equations on ${\cal M}$
(since hyper-Kahler metrics automatically obey the Einstein equations and
are self-dual)
and would be $SU(2)_N\times U(1)$ invariant (since perturbation theory
has this symmetry).  Imposing the $U(1)$ (which acts by translation of
$\sigma$, the dual of the photon) 
gives a dimensional reduction of the Einstein equations to
three-dimensional scalar-Maxwell equations on $\R^3$, with $SU(2)_N$
acting by rotations.  The only rotationally-invariant 
mode of the Maxwell field in three dimensions is the 
``magnetic charge,'' the integer $s$ that we already encountered at one loop.
The $s$-wave mode of the scalar is related by self-duality of the metric
to the ``magnetic charge'' so is likewise determined at one loop.
Thus, the whole perturbation series is determined by the one-loop term
plus the equations of hyper-Kahler geometry.

As in four dimensions, however, there can be instanton corrections
to the metric, the relevant instantons here being BPS monopoles.
For $N_f=0$, it is clear that instantons contribute to the metric.
In fact, the non-derivative $\psi\psi\psi\psi e^{-(I+i\sigma)}$ vertex
described above is part of the supersymmetric completion of a correction
to the metric.  So there is a one-instanton contribution to the
metric for $N_f=0$.  What happens for $N_f>0$?  There will be hypermultiplet
zero modes in a monopole field, so that the one-instanton field
gives a vertex $\psi^4\chi^{2N_f}e^{-(I+i\sigma)}$ ($\chi$ being
fermion components of the hypermultiplet, of opposite $U(1)_N$ 
charge {}from
$\psi$), which has too many fermions to be related by supersymmetry to
the metric on ${\cal M}$.  A correction to the metric still
might arise {}from  an $r$-instanton
contribution with $r>1$.  
Since the $U(1)_N$ charge carried by vector or hypermultiplet zero modes
could be determined {}from an index theorem and is proportional to $r$,
an $r$-instanton contribution will give in the first instance
a vertex $\psi^{4r}\chi^{2rN_f}e^{-r(I+i\sigma)}$.  
However, in integrating over bosonic collective coordinates
and computing various quantum corrections,
$\psi$ and $\chi$  zero modes of opposite charge
 might pair up and be lifted.  This process  might generate
 a vertex $\psi^4e^{-r(I+i\sigma)}$ -- which would be related
by supersymmetry to a correction to the metric -- if $2rN_f=4r-4$ or in
other words
\eqn\bombo{ r={1\over 1-N_f/2}.}
But we also need $r$ to be a positive integer, since BPS monopoles
only exist for such values of $r$. (Considering anti-monopoles instead
of monopoles reverses all quantum numbers and leads to the same restriction
on $r$; in fact, since the metric is real, there is an anti-monopole
contribution if and only if there is a monopole contribution.)   
So the only cases are $N_f=0$ and $r=1$, or $N_f=1$ and $r=2$.\foot{
For $N_f>0$, there is a symmetry reason
that only even $r$ can contribute to the metric.  The relevant
symmetry is the one that changes the sign of just one of the  
half-hypermultiplets
(and so extends $SO(2N_f)$ to $O(2N_f)$).  Since in a one-monopole field
the fermion zero mode measure is odd under this symmetry,
the symmetry must be defined to shift $\sigma$ by
$\pi $.  The $\chi$ zero modes are odd under this $\Z_2$ for
odd $r$ and $N_f>0$, implying that they cannot be lifted.}
(The fact that only one value of $r$ appears we take to mean that the exact
metric is determined by this one contribution together with the 
non-linear Einstein equations.)

In sum, then, for $N_f=0$ we expect a metric with a 
perturbative contribution
that gives $s=-4$, plus monopole corrections, and for $N_f=1$ we expect
a metric with a perturbative correction that gives $s=-2$, plus monopole
corrections.  For $N_f=2$, the perturbative and monopole corrections 
both vanish, and the quantum metric should very plausibly coincide
with the classical metric, that is, the flat metric on 
$(\R^3\times \S^1)/\Z_2$.  For $N_f>2$, there is a perturbative correction
at infinity, with $s=2N_f-4$, and the  monopole corrections vanish.

\bigskip\noindent
{\it String Theory And  Field Theory}

Let us now recall  the expectations {}from string theory
\seiberg:

(1) For $N_f=0,1$, the metric on moduli space is expected to be complete
and smooth.

(2) For $N_f\geq 2$, one expects the metric to have a $D_{N_f}$ singularity.

To clarify the meaning of the second statement, 
recall that for $N_f>2$, the $D_{N_f}$ singularity is the singularity
obtained by dividing $\C^2$ by the dihedral group
$\Gamma_{N_f-2}$.  This group was introduced earlier and is generated
by elements $\alpha,\beta$ with $\alpha^2=\beta^{N_f-2}=-1$ (the
symbol $-1$ simply denotes a central element of the group),
and $\alpha\beta=\beta^{-1}\alpha$.  For $N_f=2$, something special
happens: $D_2$ is the same as $A_1\times A_1$, or $SU(2)\times SU(2)$,
so a $D_2$ singularity should be simply a pair of $A_1$ singularities,
that is, $\Z_2$ orbifold singularities.

Let us now make
a preliminary comparison of the string theory statements with
what we have learned {}from field theory. For $N_f>2$ we have found
that {\it topologically} the moduli space ${\cal M}$ looks near 
infinity like
$\C^2/\Gamma_{N_f-2}$.  (The {\it metric} near infinity on ${\cal M}$
does not look like the obvious flat metric on $\C^2/\Gamma_{N_f-2}$.)  
We actually want to express the singularity near the origin rather
than the behavior at infinity in terms of $\Gamma_{N_f-2}$; we will do
this momentarily.
Likewise, for $N_f=2$, the moduli space that we claim,
namely $(\R^3\times\S^1)/\Z_2$, indeed has a pair of $\Z_2$ orbifold
singularities ({}from the two $\Z_2$ fixed points on $\R^3\times\S^1$)
as expected.

For $N_f=2$, a 
more precise comparison of the string theory and field theory
results is possible.  In fact, {}from string theory one can see why
the moduli space should be $(\R^3\times \S^1)/\Z_2$ with the flat
metric, just as we have found {}from field theory.  
There are many possible approaches to this result, but a quick
way is to compactify $M$-theory on $\R^7\times {\rm K3}$ and consider a 
two-brane whose world-volume fills out $\R^3\times \{p\}$, where
$\R^3$ is a linear subspace of $\R^7$ and $p$ is a point in K3.
Consider the quantum field theory on the world-volume of this
two-brane. The moduli space of vacua of this theory is the K3 manifold
itself, which parametrizes the choice of $p$.  
By arguments as in \seiberg, in various limits in 
which heavier modes decouple, this theory will reduce at low
energy to the three-dimensional $N=4$ super Yang-Mills theory
with gauge group $SU(2)$.  In particular, in K3 moduli space, 
there is a locus in which the K3 looks like $(\T^3\times\S^1)/\Z_2$
with the flat metric.  Taking the $\T^3 $ to be large and restricting
to a neighborhood of a $\Z_2$ fixed point in $\T^3$, one gets
a piece of the K3 that looks like a flat $(\R^3\times\S^1)/\Z_2$. In 
this piece of the K3, there are two $A_1$ singularities, giving on 
$\R^7$ a gauge symmetry  $SU(2)\times SU(2)=SO(4)$, which will
be observed as a global symmetry along the two-brane world-volume.
The global symmetry means that the world-volume theory is the
$N_f=2$ theory, and by construction its moduli space is 
$(\R^3\times \S^1)/\Z_2$ with flat metric, as was claimed above.
\def\N{{\cal N}}

\bigskip\noindent
{\it Comparison To Exact Metrics}

To learn more, we compare now  to what is known \ah\ about  
four-dimensional
hyper-Kahler manifolds with an $SO(3)$ symmetry of the appropriate kind.
Assuming that one wants a metric with at most isolated singularities,
the possibilities are extremely limited.  For a smooth manifold
with these properties, there are only two possibilities.  
One (sometimes  called the Atiyah-Hitchin manifold;
it was studied in \ah\ because of its interpretation as the
two-monopole moduli space) is a complete
hyper-Kahler manifold $\N$, with fundamental group $\Z_2$.  Topologically,
$\N$ looks like a two-plane bundle over ${\bf RP}^2$.  The structure
at infinity looks like $L_{-4}/\Z_2$, corresponding to a one-loop
correction with $s=-4$.  The other possibility, 
which we will call $\bar \N$, is the simply-connected 
double cover of $\N$; it is topologically a two-plane
bundle over $\S^2$, 
and the structure at infinity looks like $L_{-2}/\Z_2$,
corresponding to a one-loop correction with $s=-2$.  
Since we found $s=-4$ and $s=-2$
for the two cases -- $N_f=0,1$ -- for which we expect a smooth
metric, we propose that the $N_f=0$ theory has moduli space  
$\N$, and the $N_f=1$ theory has moduli space $\bar\N$.\foot{The extra
$\Z_2$ symmetry of $\bar \N$ which we mod out by to get $\N$ is
the  global symmetry of the microscopic $N_f=1$ theory, mentioned
earlier, that prevents a one-monopole correction to the metric
for $N_f=1$. That this symmetry acts freely on the moduli
space -- even in the strong coupling region -- is related
to the discussion of confinement that we give later.} We will discuss
in more detail 
the fundamental group and its physical interpretation later.

Now let us discuss the possible singular metrics.  According to \ah,
a hyper-Kahler metric with the requisite sort of symmetry and only
isolated singularities is severely constrained.  Such a manifold is
topologically $\C^2/\Gamma$, where $\Gamma$ is 
a cyclic or dihedral subgroup of $SU(2)$ (or if the metric is flat,
$\Gamma$ may be any finite subgroup).  
As for the metric on $\C^2/\Gamma$,  it may
be flat, but there is a more general possibility. As 
the space at infinity looks like $\S^3/\Gamma$, which is an $\S^1$
bundle over $\S^2$ or ${\bf RP}^2$, one can have a metric -- a variant
of the Taub-NUT metric -- in which
the $\S^1$ approaches at infinity an arbitrary radius $R$.\foot{
There are a few subtleties here relative to
assertions  in \ah\ that  reflect the fact that
the authors of \ah\ wanted smooth metrics with an $SO(3)_N$ action,
rather than $SU(2)_N$.  They therefore construct the Taub-NUT metric
with a $\Z_2$ orbifold singularity, and do not make explicit that it
has a smooth double cover (acted on by $SU(2)_N$ instead of $SO(3)_N$)
that can be divided by any cyclic or dihedral group $\Gamma$ (the quotient
is acted on by $SO(3)_N$ except in the case that $\Gamma$ is cyclic
of odd order).  We here need these slight generalizations.}
$R$ can be varied
simply by multiplying the metric by a constant; the  flat
metric on $\C^2/\Gamma$ is obtained in the $R\to\infty$ limit.    In the
present problem, we want $R$ of order $e$, since that is the circumference
of the circle obtained by dualizing the photon. 

Given that the $SU(2)$ gauge theory with $N_f>2$ hypermultiplets
has moduli space $\C^2/\Gamma$ for some $\Gamma$, all that
really remains is to identify $\Gamma$.
But we have 
determined that at infinity the structure 
looks like $\S^3/\Gamma_{N_f-2}$,
so  $\Gamma=\Gamma_{N_f-2}$.  Hence the moduli space has
a $\Gamma_{N_f-2}$  orbifold singularity  at the origin.  Since, in the
association of subgroups of $SU(2)$ with $A-D-E$ groups, $SO(2N_f)=D_f$
corresponds to $\Gamma_{N_f-2}$, we have confirmed {}from field theory
the string theory claim \seiberg\ that the theory with $N_f$ hypermultiplets
has a $D_{N_f}$ singularity.

It is easy to consider $U(1)$ gauge theories in a similar way.  We saw
that the $U(1)$ gauge theory with $M$ charge one hypermultiplets has a
one-loop correction with $s=M$, and that the moduli space at infinity
looks like $\C^2/\Z_M$.  Hence in this case, $\Gamma=Z_M$, and there
is a $\Z_M$ orbifold singularity at the origin.  This confirms the
claim \seiberg\ that the $U(1)$ theory with $M$ charge one
hypermultiplets has a $\Z_M$ (or $A_{M-1}$) singularity in the strong
coupling region.  For $M=1$, this means that the moduli space is completely
smooth.   The metric for $M=1$ 
is uniquely determined by the symmetries, smoothness,  and asymptotic
behavior to be the smooth Taub-NUT metric.

The Taub-NUT-like metrics on $\C^2/\Gamma$ have a very simple
structure.  They are given by an elementary closed formula (\ah,
p. 76).  In fact, in addition to the $SO(3)$ symmetry, the Taub-NUT
metrics have an extra $U(1)$ symmetry that acts by translation of the
scalar $\sigma$ which is dual to the photon; this is a precise
statement of the absence of monopole corrections.  On the other hand,
the metric on $\N$ or its double cover, while exponentially close to a
Taub-NUT type metric at infinity, has (\ah, p. 77) exponentially small
corrections which violate the extra $U(1)$ and which we interpret as
monopole corrections.

It may seem somewhat odd that the metric for $N_f>1$ is so different
{}from what it is for $N_f\leq 1$.  It is perhaps comforting, therefore,
that (\ah, p. 56) in a sense, the manifold $\bar \N$ is a kind
of analytic continuation of the $D_{N_f}$ space to $N_f=1$.
In fact, as a complex manifold, the Taub-NUT space for $D_{N_f}$ is
described by the equation
\eqn\poolo{y^2=x^2v-v^{N_f-1}.}
This has a $D_{N_f}$ singularity at $y=x=v=0$, for $N_f\geq 2$, and
two $A_1$ singularities (at $y=v=0$, $x=\pm 1$) for $N_f=2$.  If one
simply sets $N_f=1$, the same formula does give the complex structure
of $\bar\N$ -- though there is no longer a singularity.  We will
return to this formula for the complex structure in section three.

\subsec{Some Physical Properties}

We will use these results to discuss some physical properties of these
models.

First we consider symmetry breaking.  For any $N_f\not=2$, on the
generic orbit $SO(3)_N$ is broken to a finite subgroup.  (For $N_f=2$,
the generic unbroken group is $O(2)$.)  What happens in the strong
coupling region?  For $N_f\geq 2$, the $SO(3)_N$ is completely
restored at the strong coupling orbifold points.  For $N_f=0,1$, this
is not so.  The most degenerate $SO(3)_N$ orbit in $\N$ is a copy of
${\bf RP}^2$; in $\bar\N$ the most degenerate orbit is a copy of
$\S^2$.  So the maximal unbroken subgroup of $SO(3)_N$ is $O(2)$ or
$SO(2)$ for $N_f=0$ and $N_f=1$.

We now turn to consider the significance of the fundamental
group of $\N$ and $\bar \N$.

The $N_f=0$ theory has no fields with half-integral gauge quantum
numbers, so it can be meaningfully probed with external charges in
such a representation.  Let us consider the fields that would be
produced by such a charge.  In terms of the photon, an external charge
produces in $2+1$ dimensions an electric field varying as $1/r$; to be
more precise, in Cartesian coordinates $x_a,\,\,a=1,2$ with
$r=\sqrt{x_1^2+x_2^2}$, the electric field is $E_a\sim x_a/r^2$.
After performing a duality transformation, the external charge becomes
a vortex for the dual scalar $\sigma$; that is, $\sigma$ jumps by
$2\pi$ in circumnavigating the external charge.  The energy of such a
vortex has a potential logarithmic infinity both at short distances
and at large distances.  The behavior at short distances should be cut
off for our present purposes, but the behavior at long distances is
physically significant; it reflects logarithmic confinement of
electric charge in weakly coupled $2+1$-dimensional QED.

To describe this situation in a more general language, we can say that
along a circle that runs around the external charge, the fields make a
loop in the moduli space ${\cal M}$ of vacua.  If this loop is trivial
in $\pi_1({\cal M})$, then even in the low energy theory one can see
that the ``vorticity'' produced by the external charge is not really
conserved, and that the external charge can be screened.  If the loop
is non-trivial in $\pi_1({\cal M})$, then the external charge cannot
be screened in the low energy theory, though it is still conceivable
that it can be screened by massive modes that have been integrated out
in deriving the low energy theory.

For $N_f=1$, the loop produced by an external charge is automatically
trivial in $\pi_1({\cal M})$ since in fact ${\cal M}=\bar\N$ is simply
connected.  This is in accord with the fact that the $N_f=1$ theory
has isospin one-half fields, so that external charges can be screened.
For $N_f>1$, in order to make this argument, one has to decide how a
low energy physicist would understand the singularities.  However, at
least for $N_f>2$ where the moduli space (being a cone $\C^2/\Gamma$)
is contractible, it is plausible that a physicist knowing only the low
energy structure would determine that the external charges can be
screened.

For $N_f=0$, however, the answer is quite different.  The loop $C$
produced by an external charge is the generator of $\pi_1(\N)$, as we
will see momentarily, so the external charge cannot be screened
either in the low energy theory or microscopically.  In showing that
$C$ is the generator of $\pi_1(\N)$, the point is that in the analysis
in chapter nine of \ah, the fundamental group at infinity in the
moduli space is generated by two circles, defined respectively by one-forms
that were called 
$\sigma_1$ and $\sigma_2$.  (Loops wrapping once around these circles
give our standard generators $\alpha$ and $\beta$ of the fundamental
group at infinity, which for $\N$ is what we called $\Gamma_2$.)
Moreover, the metric was described in terms of functions $a,b,$ and
$c$.  Since at infinity in the moduli space, $b$ approaches a limit
and and $a$ and $c$ diverge, it is the circle defined by $\sigma_2$
that corresponds in the semi-classical region to the photon and so to
the loop $C$.  On the other hand, on the exceptional ${\bf RP}^2$
orbit, $a=0$ and $b\not=0$.  Hence the $\sigma_1$ circle can be
contracted in the interior of $\N$, and the $\sigma_2$ circle -- that
is the loop produced by the external charge -- survives as the
generator of $\pi_1(\N)$, as we wanted to show.\foot{This also means that
the $\Z_2$ symmetry of $\bar\N$, by which one would divide to get $\N$,
 is a $\pi$ shift of the scalar $\sigma$ dual
to the photon; as explained in connection with \bombo, this symmetry
must be accompanied by a sign change of one half-hypermultiplet.}

One might ask, for $N_f=0$, what sort of confinement is observed in
this theory.  As long as the vacua parametrized by $\N$ are precisely
degenerate, the energy of a pair of external charges separated a
distance $\rho$ will grow only as $\log \rho$, since the energy of a
vortex configuration of massless fields has only a logarithmic
divergence in the infrared; such a vortex configuration will form
between the two external charges.  However, suppose that one makes a
generic small perturbation of the $N_f=0$ theory that lifts enough of
the vacuum degeneracy so that a loop that generates $\pi_1(\N)$ cannot
be deformed into the space of exact minima of the energy.  (It does
not matter whether the perturbation preserves some supersymmetry.)
Then the fields on a contour that encloses one external charge but not
the other cannot be everywhere at values that exactly minimize the
energy.  In such a situation, a sort of string will form between the
external charges (one might think of it as a domain wall ending on
them), and the energy will grow linearly in $\rho$.  Thus, like the
four-dimensional $N=2$ theory \sw\ with $N_f=0$, the three-dimensional
$N=4$ theory with $N_f=0$ does not have linear confinement but gives
linear confinement after a generic small perturbation.

Finally, note that the association here of confinement with
$\pi_1(\N)$ is somewhat analogous to the association in some
four-dimensional $SO(N)$ gauge theories of confinement with
$\pi_2$ of a moduli space 
\ref\oldwitten{E. Witten, ``Global Aspects
Of Current Algebra,'' Nucl. Phys. {\bf B223} (1983) 422.}.

\nref\barsol{T.H.R. Skyrme, Proc.Roy.Soc. {\bf A260} (1961) 127; 
E. Witten, Nucl. Phys. {\bf 223} (1983) 433.}%

Another issue of physical interest stems from $\pi_2$ of the moduli
spaces.  Since these groups are non-trivial, the low energy theory on the
moduli spaces can have solitons.  There is no reason to expect these
solitons to be BPS-saturated at the generic vacuum on the moduli
space.  Furthermore, their detailed properties can depend on higher
dimension operators which are not considered in this paper.
Nevertheless, the topology of the moduli spaces supports solitons
which are localized excitations in the three-dimensional theory.  Their 
interest
is related to the fact that most
of the global symmetry of the theory does not act on the Coulomb
branch of the moduli space.  For example, all the light fields are
invariant under the global $SU(N_f)$ symmetry of the $U(1)$ gauge
theories or the $SO(2N_f)$ of the $SU(2)$ gauge theories.  We claim
that these solitons are in the adjoint representations of these
groups.  This is easiest to establish using the string theory
viewpoint \refs{\sen - \seiberg}.  The $M$-theory two-brane can wrap
non-trivial two cycles to yield zero-branes which are $SU(N_f)$ or
$SO(2N_f)$ gauge bosons.  Our solitons can be interpreted as bound
states of such a gauge boson with a two brane at a generic point in
its moduli space.  From a three-dimensional viewpoint, these solitons
are bound states of the elementary hypermultiplets.  They are bound by
the logarithmic Coulomb forces to neutral composites.  This situation
is similar to current algebra in four dimensions.  There, the
non-trivial $\pi_3$ of the moduli space leads to solitons.  Their
topological charge is identified
\refs{\barsol,\oldwitten} with the global $U(1)$ baryon number, 
which exists in the microscopic theory.
In both situations the global symmetry of the microscopic theory
manifests itself through the topology of the moduli space.

\subsec{Incorporation Of Bare Masses}

We will now try to discuss the incorporation of bare masses for
the hypermultiplets.

In four dimensions, the bare mass of a hypermultiplet is a complex
parameter, with two real components, while in three dimensions
a third parameter appears.  This arises as follows.
In four dimensions, the group that we have called $SO(3)_N$ is reduced
to an $SO(2)$ group, usually called $U(1)_R$.  A complex hypermultiplet
mass parameter carries $U(1)_R$ charge, or equivalently,
its real and imaginary parts 
transform as a vector of  $SO(2)$.  In three dimensions, as the
$SO(2)$ is extended to $SO(3)_N$,  the mass vector gets a third
component to fill out the vector representation of $SO(3)_N$.
It is easy to reach the same conclusion by viewing the masses as
expectation values of background fields in vector multiplets that gauge
some of the flavor symmetries
\ref\aps{P.C. Argyres, M.R. Plesser and N. Seiberg,
``The Moduli Space of Vacua of $N=2$ SUSY QCD and Duality in $N=1$ SUSY
QCD,'' hep-th/9603042.}.
Since all the bosons in the vector multiplets originate from gauge fields
in six dimensions
and since the masses are scalars in three dimensions, they must be
in a vector representation of $SO(3)_N$.  This interpretation also
makes it obvious that they are in the adjoint representation of the flavor 
group
$SO(2N_f)$ ($SU(N_f)$ in the $U(1)$ gauge theory).  Requiring that the 
background
fields should preserve supersymmetry means that they can all be gauged to a 
common
maximal
torus of the flavor group, and this is why there are precisely $N_f$ triplets
of mass parameters. 

In general, in four-dimensional $N=2$ theories, the moduli space of
the Coulomb branch of vacua parametrizes \sw\ a family of complex
tori.  The total space of the family is a complex manifold ${\cal M}'$
with a holomorphic two-form $\omega$, and, according to section 17 of
\secondsw, the dependence of ${\cal M}'$ on the masses is determined
by the requirement that the periods of $\omega$ vary linearly in the
masses.

The moduli space ${\cal M}$ of vacua in three dimensions is a
hyper-Kahler manifold which in fact is the analog of ${\cal M}'$; this
relation will be elucidated in the next section.  The analogs of
$\omega$ are the three covariantly constant two-forms $\omega_a$,
$a=1,2,3$ of the hyper-Kahler manifold ${\cal M}$ (two of which
correspond to the real and imaginary parts of $\omega$).  These
transform in the vector representation of $SO(3)_N$.  We normalize
them in the semi-classical region of large $\phi$ to be independent of
the hypermultiplet bare masses.

The natural three-dimensional analog of the four-dimensional statement
that the periods of $\omega$ vary linearly in the masses is then a
three-dimensional statement that the periods of the $\omega_a$ should
vary linearly with the masses.  Notice that such an assertion is
compatible with $SO(3)_N$, as both the mass parameters and the
two-forms transform as $SO(3)_N$ vectors.  A direct field theory
justification of this principle in three-dimensional $N=4$ models is
not clear at the moment.
\foot{But note that for those three-dimensional $N=4$ models that have
been related to string theory \seiberg, which include those studied in
detail in this paper, the fact that the periods of $\omega_a$ vary
linearly in the masses follows {}from the fact that the periods of the
$\omega_a$ are the natural coordinates parametrizing hyper-Kahler
metrics on K3.}  We will here simply accept this principle and discuss
its implementation for the $SU(2)$ theory with $N_f$ doublets.

First we consider the case $N_f=0$.  The moduli space $\N$ that we
proposed is homotopic to the two-manifold ${\bf RP}^2$.  As this is
unorientable, the two-dimensional homology of this manifold has rank
zero, and a closed two-form has no periods.  Thus, there is no way to
perturb this model to include mass parameters.  That is just as well,
since no hypermultiplets are present in the model.

Now consider $N_f=1$.  The moduli space $\bar\N$ is homotopic to
$\S^2$; a closed two-form on this manifold has a single period, the integral
over $\S^2$.  Thus, a single
``mass vector'' can be introduced, compatible with the fact that the
model has $N_f=1$.  In fact, the hyper-Kahler metric that is the
appropriate deformation of $\bar \N$ to include masses has been
described explicitly by Dancer
\ref\dancer{A. S. Dancer, 
``Nahm's Equations And Hyperkahler Geometry,'' 
Commun. Math. Phys. {\bf 158}
(1993) 545, ``A Family Of Hyperkahler Manifolds,'' preprint.}.
Dancer constructs a deformation $\bar\N_{\vec \lambda}$ of the
hyper-Kahler manifold $\N$ depending on an $SO(3)_N$ vector 
$\vec\lambda$.
That the periods of $\vec\omega$ vary linearly with $\vec\lambda$ is
a consequence of Dancer's construction of $\bar\N_{\vec\lambda}$ as
a $U(1)$ hyper-Kahler quotient (of a hyper-Kahler eight-manifold) with
$\vec\lambda$ as the constant term in the moment map.  We will return
to Dancer's manifold in section three.

For $N_f>1$, the real homology of the resolution of the $D_{N_f}$
singularity is known to have two-dimensional homology of rank $N_f$,
so that $N_f$ mass vectors can be introduced.

It is now by the way clear, even without solving for hyper-Kahler metrics
as in \ah, that for $N_f>2$ the metric on the moduli space of vacua
must be singular.  An $SO(3)$ action with three-dimensional orbits on
a four-manifold constrains the topology so much that there could not be
$N_f>2$ independent two-cycles, unless some or all are collapsed at a
singularity.

Even though we have not determined the metric, it is easy to see how
the masses affect the singularity of the moduli space.  First, physically,
we expect that if only $k<N_f$ masses vanish the singularity should
be $D_k$.  Furthermore, if $n$ masses are equal and non-zero we expect an 
$A_{n-1}$
singularity (classically, upon adjusting the Higgs field to
cancel the bare mass of some of the fields,   we get a $U(1)$ gauge theory
with $n$ massless hypermultiplets, which gives an $A_{n-1}$ singularity, from
which a Higgs branch emanates).
This is exactly the behavior after the $D_{N_f}$ singularity
is blown up.  The $N_f$ mass parameters are the parameters labeling the
blow-up of the singularity.

\newsec{Field Theory On $\R^3\times \S^1_R$}

In the remainder of this paper, we will mainly be studying four-dimensional
$N=2$ super Yang-Mills theory formulated on a space-time $\R^3\times \S^1_R$,
where $\S^1_R$ is a circle of circumference $2\pi R$.  We focus on the case
of gauge group $G=SU(2)$, with $N_f\leq 4$ matter hypermultiplets in the
two-dimensional representation.  (The upper bound on $N_f$, which has no
analog in three dimensions, ensures a non-positive beta function in the 
four-dimensional theory.)  We recall that the bosonic fields of the theory
are the $SU(2)$ gauge field and a complex scalar $\phi$ in the adjoint
representation.

To begin with, we consider what happens for $R$ much greater than the
natural length scale of the four-dimensional theory (which is set by
an appropriate bare mass, order parameter, or by the scale parameter
$\Lambda$ introduced in quantizing the theory).  In this regime, one
can borrow four-dimensional results.  The moduli space of vacua in
four dimensions is \refs{\sw, \secondsw}\ the complex $u$ plane, where
$u=\Tr\,\phi^2$ is the natural order parameter.  The massless bosons
are $u$ and an Abelian photon, which we will call $A$.  The effective
action for $A$, in four dimensions, looks like
\eqn\rmo{L=\int d^4x\left({1\over 4e^2}F_{\mu\nu}F^{\mu\nu}+{i\theta
\over 32 \pi^2}F_{\mu\nu}\tilde F^{\mu\nu}\right).}
Here $\mu,\nu =1\dots 4$ are space-time indices, $F_{\mu\nu}=\partial_\mu
A_\nu-\partial_\nu A_\mu$, and $\tilde F_{\mu\nu}=\half\epsilon_{\mu\nu
\alpha\beta}F^{\alpha\beta}$.  $e$ and $\theta$ are functions of $u$
and were determined in \refs{\sw,\secondsw}.  
A key point in computing them was to interpret
$e$ and $\theta$ as determining the complex structure of an elliptic curve
$E$.  The most natural convention in defining $E$, in the case $N_f\not= 0$,
was explained on pp. 487-8 of \secondsw.  $E$ is the complex torus
with $\tau$ parameter
\eqn\taupar{\tau={\theta\over \pi}+{8\pi i\over e^2}.}
$E$ is isomorphic in other words to $\C/\Gamma$, where $\Gamma$ is the 
lattice in the complex plane generated by the complex numbers 1 and $\tau$.
For $N_f=0$, one can also conveniently use, as in \sw, an isogenous
elliptic curve with $\tau$ replaced by $\tau/2$, but this is awkward if
one wishes to let $N_f$ vary.

Once we work on $\R^3\times \S^1_R$, there is a small subtlety about
defining the theory in the $N_f=0$ case.  In quantizing a gauge
theory, one must divide by the group of gauge transformations. But
precisely what gauge transformations do we want to divide by?  Do we
want to consider gauge transformations which, in going around the
$\S^1_R$, are single-valued in $SU(2)$, or gauge transformations that
would be single-valued only in $SO(3)$?  For $N_f \not= 0$, since
there are fields that are not invariant under the center of $SU(2)$,
the gauge transformations that are single-valued only in $SO(3)$ are
not symmetries, so one is forced to divide only by the smaller group.
For $N_f=0$, one is free to divide by either the larger or the smaller
group; the two choices give slightly different (but obviously closely
related) quantum theories, the moduli space of vacua of one being a
double cover of the moduli space of vacua of the other.  To obtain
results that vary smoothly with $N_f$, in quantizing the $N_f=0$
theory, we will divide only by the ``small'' gauge group, the gauge
transformations that are single-valued in $SU(2)$.  It will be seen,
as one might expect, that this choice will agree with \taupar, while
the other choice has the effect of replacing $\tau$ by $\tau/2$.

To determine what happens in compactification on $\R^3\times \S^1_R$ for
very large $R$, we simply expand \rmo\
in terms of fields that are massless in the three-dimensional sense.
These are the fourth component $A_4$ of the gauge field and also a 
three-dimensional photon $A_i$, $i=1,\dots,3$ which is dual to another
scalar $\sigma$.  First of all, the gauge field $A$ in \rmo\ is
normalized (see the discussion of eqn. (3.12) in \sw) so that fields in
the two-dimensional representation of $SU(2)$ have half-integral charges,
and the magnetic flux of a magnetic monopole is $4\pi$.  Because of the
first assertion, and the fact that we are only dividing by the gauge 
transformations that are single-valued in $SU(2)$, $\int_{\S^1_R}A$ is
gauge-invariant modulo $4\pi$.  We therefore write the massless
scalar coming {}from $A_4$ as 
\eqn\qko{A_4={b\over \pi R},}
where $b$ is an angular variable, $0\leq b\leq 2\pi$.

The effective action becomes in terms of $b$ and the three-dimensional
photon
\eqn\quicko{L=\int d^3x \left({1\over \pi Re^2}|db|^2+{\pi R\over 2e^2}
F_{ij}^2+{i\theta\over 8\pi^2}\epsilon^{ijk}F_{ij}\partial_kb\right).}
The next issue is to dualize the three-dimensional photon.  To do so,
introduce a two-form $B_{ij}$ with (in addition to standard gauge
invariance $A_i\to A_i+\partial_iw$) an extended gauge-invariance
\eqn\ugg{A_i\to A_i+C_i,~~ 
B_{ij}\to B_{ij}+\partial_iC_j-\partial_jC_i}
where $C$ is an arbitrary connection on a line bundle,
and introduce also a scalar field $\sigma$ with $0\leq \sigma\
\leq 2\pi$.  Replace the $F$-dependent part of \quicko\ by
\eqn\kormo{\int d^3x\left(
{\pi R\over 2e^2}(F_{ij}-B_{ij})^2+
{i\theta\over 8\pi^2}\epsilon^{ijk}(F_{ij}-B_{ij})
\partial_kb + {i\over 8\pi} \epsilon^{ijk}B_{ij}\partial_k\sigma\right).}
The point of this is that if one first integrates over $\sigma$, then
$\sigma$ serves as a Lagrange multiplier, enabling one to set
$B=0$ modulo an extended gauge transformation \ugg; in this
way one reduces \kormo\ to the relevant part of \quicko.  On the
other hand, one can use the extended gauge invariance \ugg\
to set $F=0$, whereupon after integrating over $B$ one gets
a dual description with a massless scalar $\sigma$.  The
dual formula for the low energy action is in fact
\eqn\jurry{\tilde L= \int d^3x\left({1\over \pi Re^2}|db|^2
+{e^2\over \pi R (8\pi)^2}
\left|d\sigma-{\theta\over\pi}db\right|^2\right).}
This is a sigma model in which the target space is a two-torus $E$ 
with the $\tau$ parameter given in \taupar.  (Had we chosen
to divide by the ``big'' group of gauge transformations,
$b$ would have been replaced by $b/2$, and $\tau$ by $\tau/2$,
giving formulas related to the other description of the $N_f=0$ theory.)

Moreover, the area $V_E$ of $E$ is
\eqn\ingo{V_E(R)= {1\over 16\pi R}.}
An overall multiplicative constant in \ingo\ depends on exactly
how one writes the effective 
action of a sigma model in terms of the metric
of the target space, but the $R$ dependence of $V_E$ is significant.
We see immediately that near four dimensions, that is for $R\to\infty$,
the torus $E$ is small.

One can likewise work out other terms in the effective action of
the theory on $\R^3\times \S^1_R$.  For instance,  on $\R^4$,
the effective action for $u$ is given by an expression
\eqn\murko{\int d^4x \,\,g_{u\bar u}du\,d\bar u,}
where $g_{u\bar u}$ is a metric on the $u$ plane computed in \sw.
After compactification on $\S^1_R$, one gets the three-dimensional
effective action, in the large $R$ approximation, simply by integrating
over $\S^1_R$, giving
\eqn\nurko{\int d^3x \,2\pi R\,g_{u\bar u}du\,d\bar u.}

\subsec{First Look At The Moduli Space}

Now we can describe the moduli space ${\cal M}$ of vacua of the $N=2$ theory
compactified on $\R^3\times \S^1_R$, at least for large $R$.
The vacua are labeled by the order parameter $u$, together with, for
every  $u$,
 an additional complex torus $E_u$.   {}From what
we have just seen, the relevant family of tori is the {\it same} family
of tori that controls the $u$ dependence of the gauge couplings in
four dimensions.  So we can immediately borrow results {}from \secondsw.
With $\tau$ normalized as in \taupar, the appropriate family of tori
is described by the algebraic equation
\eqn\torfam{y^2=x^3-x^2u+x.}
Therefore, the moduli space of the three-dimensional theory, for large
$R$, is given by \torfam.

Actually, there are a few imprecisions here.  A minor one is that
the  equation \torfam, for
given $u$, does not describe a compact torus; one point on the torus is
at $x=y=\infty$.  This was not very important in the four-dimensional
story, where only the complex structure of $E_u$, which can still
be detected even if a point is projected to infinity, was of interest.
But after compactification to three dimensions, every point on $E_u$,
including the point $x=y=\infty$,  is
an observable vacuum state of the theory.  So if we want to be more
precise, we should extend $x$ and $y$ to a set of homogeneous coordinates
$x,y$, and $z$, and write the equation for $E_u$ in its homogeneous
form:
\eqn\homform{zy^2=x^3-zx^2u+z^2x.}
We will omit this except when it is essential.

A more far-reaching point is that while in four-dimensions it suffices
to describe $x-y-u$ space as a complex manifold (since the complex
structure of $E_u$ is all that one really needs), once one is in three
dimensions, the moduli space ${\cal M}$ has a {\it hyper-Kahler
metric}, and merely describing it as a complex manifold, as in
\homform, does not suffice.  We must complete the description by
finding the metric.  We know the large $R$ limit of the hyper-Kahler
metric, {}from \jurry\ and
\nurko.
Let us examine some aspects of that result with the aim of giving 
a formulation that makes sense for arbitrary $R$.

Note that, as the $R$ dependence of \nurko\ is inverse to that of 
\jurry, {\it the volume form on the moduli space of three-dimensional
vacua is independent of $R$}, at least in the approximation of
dimensional reduction {}from four dimensions.  That volume form is in
fact a constant multiple of $db\wedge d\sigma\wedge g_{u\bar u}
du\wedge d\bar u$.  This can be put in a more convenient form as
follows.  The differential form $dx/y$ is invariant under translations
on $E$, so it is a linear combination of $db$ and $d\sigma$, with
$u$-dependent coefficients.  Hence $|dx/y|^2=db\wedge d\sigma \cdot
f(u, \bar u)$, for some function of $u$.  But in fact $f(u, \bar
u)=g_{u\bar u }$.  For this, recall {}from \sw\ that
\eqn\ijj{g_{u\bar u}=2{\rm Im}\left({da\over du}{d\bar a_D\over d\bar 
u}\right)}
where $da/du$ and $da_D/du$ are the periods of $dx/y$.
On the other hand, {}from the Riemann relations
\eqn\jj{\int_E|dx/y|^2=2{\rm Im}\left({da\over du}{d\bar a_D\over d\bar u}
\right).} 
The conclusion, then, is that in terms of the holomorphic two-form
\eqn\hij{\omega={dx\wedge du\over y} }
on ${\cal M}$, the volume form, at least for large $R$, is just
\eqn\pij{\Theta=\omega\wedge\bar\omega.}

\subsec{$R$ Dependence Of The Metric}

Let us now go back to four dimensions as a starting point, and ask,
{}from that point of view, what happens to the dynamics of the $N=2$
theory when one compactifies {}from $\R^4$ to $\R^3\times \S^1_R$?
One still has ordinary, localized four-dimensional instantons.  The
main novelty is that one has in addition a new kind of instanton,
namely a magnetic monopole (or a dyon) that wraps around $\S^1_R$.
The action of such an instanton, for large $R$, is $I=2\pi RM$, where
$M$ is the mass of the monopole in the four-dimensional sense.

The moduli space ${\cal M}$ of vacua is a hyper-Kahler manifold.  In
one of its complex structures, the one exhibited in \homform, ${\cal
M}$ is elliptically fibered over the complex $u$ plane.  Let us call
this the distinguished complex structure.

In the distinguished complex structure, $M$ is not a holomorphic
function (rather, it is the absolute value of the holomorphic function
$a_D+na$ where $n$ is the dyon charge).  
Therefore, it is impossible for monopoles to correct the
distinguished complex structure of the moduli space.  However,
monopoles do contribute to the metric on ${\cal M}$.  In fact, for
$R=0$ these contributions were discussed in the last section, and the
case $R\not= 0$ can be treated similarly.\foot{In section two, we
found in three dimensions that there were no monopole contributions
for $N_f>1$, but this depended on a symmetry that is absent at $R>0$.}
Changing the metric on ${\cal M}$ without changing the distinguished
complex structure means that the other complex structures on ${\cal
M}$ will change.

So far, we have just given a heuristic reason in terms of monopoles
that the distinguished complex structure of ${\cal M}$ is independent
of $R$.  Two more fundamental reasons for this can be given. (1)
Picking the distinguished complex structure selects an $N=1$
subalgebra of $N=2$ supersymmetry.  This $N=1$ algebra relates $R$ to
a three-dimensional vector that comes {}from the components $g_{i4}$,
$i=1,\dots ,3$ of the space-time metric tensor $g$; that vector is
dual to a scalar $\eta$.  $N=1$ supersymmetry would require the
complex structure of ${\cal M}$ to depend on $\eta$ if it depends on
$R$, but the zero mode of $\eta$ decouples in flat space quantum field
theory. (It might not decouple in the field of a gravitational
instanton!)  (2) The string theory approach \refs{\bds,
\seiberg}, as we will explain in section      four, makes it clear that
there is $R$ dependence in the Kahler metric of ${\cal M}$ but not in the
distinguished complex structure.

There is actually a natural rationale for a change in the metric of
${\cal M}$ due to monopoles.  The complex manifold ${\cal M}$ is
smooth for $N_f=0$ as one can verify {}from \homform.  But, as was
discussed in \beckers\ in a related context, the metric obtained by
dimensional reduction as in \jurry\ and \nurko\ is not smooth; there
are singularities at points where the fiber $E_u$ has a singularity.
Those are points at which the monopole mass goes to zero and monopole
corrections cannot be ignored; it was proposed in \beckers\ that the
effect of the monopole corrections would be to eliminate the
singularities and produce a smooth hyper-Kahler metric.  For $N_f\geq
2$, ${\cal M}$ has orbifold singularities in its complex structure, as
we will review below; in that case, one would propose that with
monopole corrections included, the hyper-Kahler metric is smooth
except for the orbifold singularities present in the complex
structure.

So at this point, we know one complex structure on ${\cal M}$, and we
need a recipe to determine the smooth hyper-Kahler metric (or
hyper-Kahler metric with orbifold singularities) for given $R$.  Yau's
theorem on existence of Ricci-flat Kahler metrics has analogs in the
non-compact case 
\ref\tian{G. Tian and S.-T. Yau, ``Complete Kahler
Manifolds With Zero Ricci Curvature, I, II'' (preprints).}.  
The basic
idea is that to determine a hyper-Kahler metric, given a complex
structure, one needs (i) the non-degenerate holomorphic two-form
$\omega$, (ii) a two-dimensional class that should be the Kahler class
of the metric, (iii) a specification of the desired behavior at
infinity.

In the present case, we propose that these data should be as follows.
(i) We take $\omega$ to be $\omega=dx\wedge du/y$, as introduced above.
We ask that the hyper-Kahler metric should have $\omega\wedge\bar\omega$
as its volume form.
(ii) We specify the Kahler class of the metric by stating that the area
of $E_u$ is (as in \ingo) $V_E(R)=1/16\pi R$ and that other periods of the 
Kahler
form, if any, are independent of $R$.
(iii) Infinity in ${\cal M}$ is the region of large $u$; we specify the
metric in this region by asking that it should reduce to what was obtained
in \jurry\ and \nurko.

We will assume that with an appropriate non-compact version of Yau's
theorem, (i), (ii), and (iii) suffice to determine a unique smooth
hyper-Kahler metric on ${\cal M}$ (or a hyper-Kahler metric with only
orbifold singularities forced by the complex structure).  The most
delicate question for physics is whether (i) and (ii), which we found
in the large $R $ limit, are actually exact statements about the
quantum field theory.  In the next section, we will use string theory
to argue that this is so, but for now we take it as a plausible
assumption.  In particular, we assume, according to (ii), that the
area of $E_u$ diverges for $R\to 0$; we will now see that this has
interesting and verifiable consequences.
 
\subsec{Comparison To Three Dimensions}

In the last subsection, a proposal was made for the description
of the hyper-Kahler moduli space ${\cal M}$ that arises in compactification
of the $N=2$ theory on $\S^1_R$, for any positive $R$.
Formally speaking, as $R\to 0$, this should go over to the purely
three-dimensional $N=4$ theory, analyzed in  section two.  Our next goal
is to make this connection.  

Since we claim that the area of $E_u$ is $1/16\pi R$, something must
diverge in the limit $R\to 0$; the $E_u$ cannot remain compact.  
We earlier exhibited the compactness of $E_u$'s for $N_f=0$ by 
writing the equation in the homogeneous form
\eqn\unnu{zy^2=x^3-zx^2u+z^2x.}
This compactness will have to disappear for $R\to 0$, if our formula
for the area is correct.

Here is another reason that the compactness must be lost. At $R=0$,
the moduli space has an $SO(3)_N$ symmetry which was extensively
discussed in section two.  Since $SO(3)_N$ rotates the complex
structures, the full $SO(3)_N$ will not be manifest once one picks a
distinguished complex structure.  However, a $U(1)$ subgroup, which
preserves the distinguished complex structure, should be visible.  In
fact, one should see a $\C^*$ that preserves the complex structure, of
which the $U(1)$ subgroup preserves the metric.  But the complex
surface \unnu\ does not have a non-trivial $\C^*$ action; such a group
would have to map each $E_u$ to another $E_{u'}$ (because the
holomorphic function $u$ would have to be constant on the image of
$E_u$) and hence to itself (since the different $E_u$ have different
$j$-invariants), but a torus $E_u$ does not have a non-trivial $\C^*$
action.  So something must be deleted in order to find the $\C^*$
action.

Suppose that we throw away the points with $z=0$.  After that we can
scale $z$ to 1 and reduce to affine coordinates $x,y$.  This gives
back the original description in which the points at $x=y=\infty$ are
omitted:
\eqn\uggu{y^2=x^3-x^2u+x.}
Let $v=x-u$, giving
\eqn\puggu{y^2=x^2v+x.}
Suddenly a $\C^*$ action, with weights $ 1,2,-2$ for $y,x,v$, is apparent.
Moreover,\foot{According to p. 20 of \ah, $\bar \N$ is the complex
surface $Y^2=X^2V+1$, and $\N$ is the
quotient by    the freely acting $\Z_2$ symmetry
$X\to -X,Y\to -Y,V\to V$.  To take the quotient, we introduce the 
$\Z_2$-invariant independent variables $x=X^2$, $y=XY$ (we need not
introduce $Y^2$ since it equals $X^2V+1=xV+1$).  In terms of $x,y$, and
$v=V$, the equation $Y^2=X^2V+1$ implies $y^2=x^2v+x$, which  then
describes $\N$.}
\puggu\ gives the complex structure of the   Atiyah-Hitchin
manifold ${\cal N}$, which we have proposed as the moduli space of the
$N_f=0$ theory in three dimensions!

Thus, we propose that what must be deleted when $R\to 0$ and the area
of $E_u\to\infty$ are simply the points $x=y=\infty$.\foot{Those
points must be deleted before one can make the change of variables
{}from $x$ and $u$ to $x$ and $v$.  In fact, in homogeneous
coordinates a similar substitution $v=x-uz$ fails to be an invertible
change of coordinates at $z=0$, where $x$ and $v$ fail to be
independent.}  We will now give many checks showing how a similar
story works for $N_f>0$.  We first consider the case of zero
hypermultiplet bare mass, and then incorporate the bare mass for
$N_f=1$.

For $N_f=1$, in affine coordinates, the result obtained in 
\secondsw\ was
\eqn\juggu{y^2=x^3-x^2u+1.}
After substituting $v=x-u$, we get
\eqn\higgu{y^2=x^2v+1,}
which has the expected $\C^*$ action with weights $0,1,-2$ for $y,x,v$.
Moreover (\ah, p. 20), \higgu\ does give the complex structure
of $\bar \N$, the double cover of the Atiyah-Hitchin manifold which was
proposed in section two as the moduli 
space of the $N_f=1$ theory in three dimensions.

For $N_f=2$, the result obtained in \secondsw\ was
\eqn\bugg{y^2=(x^2-1)(x-u).}
After the substitution $v=x-u$, we get
\eqn\nugg{y^2=(x^2-1)v,}
with the expected $\C^*$ action (weights $1,0,2$ for $y,x,v$) and the
two $A_1$ singularities (at $y=v=0$, $x=\pm 1$)
expected for the three-dimensional $N_f=2$ theory.

For $N_f=3$, the result of \secondsw\ was 
\eqn\rugg{y^2=x^2(x-u)+(x-u)^2.}
The substitution $v=x-u$ gives
\eqn\kugg{y^2=x^2v+v^2,}
which is a standard form of the $A_3$ or equivalently $D_3 $ singularity,
as expected.  

Finally, for the $N_f=4$ theory with zero bare mass, one has
\eqn\polyugg{y^2=(x-e_1u)(x-e_2u)(x-e_3u).}
After linear transformations of $x$ and $u$ (that is replacing $x$ and
$u$ by certain linear combinations that will be called $x$ and $v$),
this can be put in the form
\eqn\hugg{y^2=x^2v+v^3,}
which is a standard form of the $D_4$ singularity.  (This $D_4$
singularity -- and the associated configuration of two-spheres after
deformation of the singularity -- is actually closely related to the
way $D_4$ triality was exhibited in section 17 of \secondsw.)

Note that all of these results depend on changes of variables --
mixing $x$ and $u$ -- that would be unnatural in four dimensions
(where $u$ is a physical field and $x$ is a somewhat mysterious
mathematical abstraction) but are natural in three dimensions where
$x$ and $u$ are on the same footing.

Finally, let us consider the $N_f=1$ theory with a bare mass $m$.
According to \refs{\sw,\secondsw}, the appropriate object in four
dimensions is described by the equation
\eqn\upplo{y^2=x^3-x^2u+2mx +1.}
We proposed at the end of section two that the three-dimensional $N_f=1$
theory should be described by Dancer's manifold, whose complex structure
(see the second paper cited in \dancer) is
\eqn\illop{y^2=x^2v+i\lambda x+1.} 
These agree after the usual change of variables $v=x-u$ and an obvious
identification of $\lambda$ and $m$.

\subsec{Soft Breaking To $N=1$}

One of the main tools in \sw\ was to consider what happens what one adds
to the theory a superpotential $\Delta W=\epsilon u$, softly breaking the $N=2$
supersymmetry to $N=1$.  The result was to produce two vacua with monopole
condensation, a mass gap, and confinement.

We now want to ask what happens if one makes the same perturbation after
compactification to three dimensions on $\S^1_R$.  {\it A priori}, because
of the mass gap in four dimensions, one should find the same two vacua
after compactification on  $\S^1_R$, at least if $R$ is big enough.

To investigate this, we look for critical points of the superpotential
\eqn\joker{W=\lambda\left(y^2-x^3+x^2u-x\right)+\epsilon u,}
where the chiral superfield $\lambda$ is introduced as a sort of Lagrange
multiplier to enforce the constraint $F=0$, 
where $F=y^2-x^3+x^2u-x$ 
is the quantity whose vanishing is the defining condition
of $E_u$.    The equations for
a critical point of $W$ are 
\eqn\oker{F={\partial F\over \partial y}={\partial F\over \partial 
x}={\partial F\over \partial u}=0.}
The equations \oker\ are conditions for a singularity of the fiber $E_u$.  
They are
precisely the conditions found in \sw\ for a vacuum state in the
presence of the $\epsilon u$ perturbation.  They have two solutions,
at $y=0$, $u=\pm 2,$ $x=u/2$. So we see that the two vacua found in
\sw\ indeed persist after compactification on $\S^1_R$.

In the limit, though, of $R\to 0$, a puzzle presents itself.  In three
dimensions, the $\Delta W=\epsilon u$ perturbation breaks $N=4$
supersymmetry to $N=2$, giving bare masses to fields that are not in
the $N=2$ vector multiplet.  But the minimal $N=2$ theory generates a
superpotential \ahw.  It is uniquely determined by the symmetries of 
the theory to be
\eqn\ahwsuper{W = e^{-\Phi}}
where $\Phi$ is an $N=2$ chiral superfield which originates by duality from
the massless vector multiplet.  The superpotential \ahwsuper\ does not have a 
stationary point and therefore the theory does not have a vacuum -- it
runs off to infinity.  How is this fact consistent with the above
construction?

To resolve this point, we should be more precise about some of the
above formulas, restoring the dependence on the four-dimensional
gauge-coupling $g_4(\mu)$ and the renormalization point $\mu$ (the
scale parameter $\Lambda$ is determined by $\Lambda^4=\mu^4
\exp(-8\pi^2/ g_4(\mu)^2)$).  In the equation $y^2-x^3+x^2u-x=0$, the
term linear in $x$ is an instanton effect.  To restore the dependence
on $g_4$, we should write
\eqn\ugguu{y^2=x^3-x^2u+x\mu^4 \exp(-8\pi^2 / g_4(\mu)^2).}
Now we introduce the three-dimensional gauge coupling,
defined classically by $1/g_3^2=R/g_4^2$.  (Corrections to that formula 
hopefully
do not matter for the qualitative remarks that we are about to make.)
In terms of $g_3$, \ugguu\ becomes
\eqn\yuggu{y^2=x^3-x^2u+\eta x,}
where $\eta=\mu^4\exp(-8\pi^2/Rg_3^2)$.  If we are going to compare to
\ahw, we should keep $g_3$ fixed as $R\to 0$; this means taking
$\eta\to 0$.  That is clear intuitively; the three-dimensional theory
does not have four-dimensional instantons, so in some sense the
instanton factor $\eta$ should be dropped as $ R\to 0$.  On the other
hand, we do not want to simply discard the linear term in \yuggu, as
this would not give the Atiyah-Hitchin manifold.  Instead we make a
change of variables $x-u=v$, $x=\eta\tilde x$, $y=\eta\tilde y$, and
get the Atiyah-Hitchin manifold
\eqn\pohj{\tilde y^2=\tilde x^2v+\tilde x.}

Now the superpotential $\Delta W= \epsilon u$ is in the new variables 
$\Delta W=\epsilon(\eta\tilde x-v)$. So, as  in \joker, we study
\eqn\jokern{W=\eta^2\lambda\left(\tilde y^2-\tilde x^2v- \tilde x\right)
+\epsilon(\eta\tilde x-v).}
Solving ${\partial W/ \partial \lambda}={\partial W/ \partial 
\tilde y}={\partial W/ \partial v}=0$ for $\lambda$, $\tilde y$ and
$v$ we find an effective superpotential for $\tilde x$
\eqn\jokernn{W_{eff}=\epsilon(\eta\tilde x + {1 \over \tilde x}).}
The critical points are at  $\tilde x= \pm \eta^{-1/2}$.  So for every
non-zero $\eta$ there are two vacua, but as $\eta\to 0$, the vacua run
away to infinity.  In fact, our analysis leads to a new derivation of
\ahwsuper\ for $\eta=0$, if we identify $e^{-\Phi} = 
{\epsilon/ \tilde x}$.

\newsec{String Theory Viewpoint}

In this concluding section, we will use the string theory viewpoint
\refs{\sen - \seiberg}
to explain some crucial points that entered in sections two and three:

(1) If one compactifies {}from four to three dimensions on $\S^1_R$,
then varying $R$ 
does not change the distinguished complex structure of ${\cal M}$,
which is the one in which ${\cal M}$ is elliptically fibered over
the complex $u$ plane.  
On the other hand, varying $R$ does change the Kahler metric of ${\cal M}$,
in such a way that the area of the fibers is a multiple of $1/R$.

(2) In three dimensions, the hypermultiplet bare masses correspond
to periods of the covariantly constant two-forms $\omega_a$ on the
moduli space.

The starting point is to consider $M$-theory compactification
on $\R^7\times {\rm K3}$.  Then one considers a two-brane whose
world-volume is $\R^3\times \{p\}$, where $\R^3$ is a signature $-++$
flat subspace of $\R^7$, and $p$ is a point in K3.  The quantum field
theory on the two-brane world-volume is a $2+1$-dimensional theory.
The moduli space of vacua of this theory is a copy of K3, since 
$p$ could be any point in K3.

On the other hand, this theory is dual to the Type I or
heterotic string compactified
on $\R^7\times \T^3$.  Under the duality, the $M$-theory two-brane corresponds
to a Type I five-brane wrapped over the $\T^3$ (to give a two-brane in
$\R^7$).  On the five-brane
world-volume there is an $SU(2)$ gauge symmetry.  Therefore, suitable
limits of this theory can look like $SU(2)$ or (in the event of some
high energy symmetry breaking) $U(1)$ gauge theories in $2+1$ dimensions.

The moduli space of $M$-theory on K3 is a product of two factors.
One, a copy of $\R^+$, parametrizes the K3 volume and corresponds
to the heterotic or Type I coupling constant.
The other factor is as follows.  The two-dimensional integral cohomology
of K3 is an even self-dual lattice of signature $(19,3)$; we denote
it as $\Gamma^{19,3}$.  The remaining factor in the $M$-theory moduli
space is the choice of a three-dimensional positive definite
real subspace $V^+$ of 
$\R^{19,3}=\Gamma^{19,3}\otimes_\Z\R$.  The choice of $V^+$ is equivalent
to a choice of the periods of the covariantly constant two-forms
$\omega^a$ in the hyper-Kahler metric on K3.  By the time one
gets to a limit in which one sees $2+1$-dimensional $SU(2)$ gauge theory,
a piece of the K3 is interpreted as the moduli space ${\cal M}$
of vacua \seiberg,
and the mass parameters correspond to periods that can be measured
in ${\cal M}$; that is the basic reason for (2) above.

As for the heterotic string on $\T^3$, it has a Narain lattice $\Gamma^{19,3}$,
and the moduli space is the space of three-dimensional positive
definite subspaces $V^+$ 
of $\R^{19,3}$, interpreted as the space of right-moving
momenta.

If we want to see {\it four}-dimensional quantum field theory on
$\R^{2,1}\times \S^1$, we should split the $\T^3$ as $\S^1\times \T^2$,
in such a way that the Wilson lines and $B$-field all live only on the
$\T^2$ factor.  Then we will see a five-brane compactified on $\T^2$
to $\R^3\times \S^1$; by tuning the moduli of the $\T^2$ appropriately,
we can get four-dimensional $SU(2)$ gauge theory on $\R^3\times\S^1$,
with various numbers of hypermultiplets.  
Splitting the $\T^3$ in the indicated fashion means splitting
the Narain lattice as $\Gamma^{19,3}=\Gamma^{1,1}\oplus \Gamma^{18,2}$,
in a way compatible with $V^+$; that is $V^+$ is the direct sum of
a one-dimensional subspace of $\R^{1,1}$ and a two-dimensional subspace of
$\R^{18,2}$.

In terms of $M$-theory on K3, this splitting can be accomplished by
specializing to K3's that are elliptically fibered (over ${\bf P}^1$)
with a section.
For such a K3, the fiber $F$ and section $S$ obey $F\cdot F=0$,
$F\cdot S=1$, $S\cdot S=-2$, and generate a $\Gamma^{1,1}$ subspace of
the cohomology.  On such a K3, there is a distinguished complex structure,
the one in which the K3 is elliptically fibered.  In any limit in which
a piece of the K3 turns into the moduli space ${\cal M}$ of a field
theory, ${\cal M}$ will inherit a distinguished complex structure
in which it is elliptically fibered, explaining part of point (1) above.

In terms of K3, the compatibility of $V^+$ with the splitting 
$\Gamma^{19,3}=\Gamma^{1,1}\oplus \Gamma^{18,2}$ means that the
Kahler form is an element of $\R^{1,1}$ (while the real and imaginary
parts of the holomorphic two-form $\omega$ lie in $\Gamma^{18,2}$).
The Kahler form is therefore dual to a linear combination of $F$ and $S$,
leaving two parameters of which one can be regarded as the overall volume
of K3, while the second is the area of the fiber $F$.  In the constructions
of \refs{\sen - \seiberg}, the volume of the K3 (or heterotic string
coupling constant) does not correspond to
an interesting modulus of the $2+1$-dimensional 
or $3+1$-dimensional field theories, so we just fix it.
The remaining moduli are then the area of $F$ (which is varied while
keeping fixed the volume) and the choice of the complex
structure of the elliptic fibration, which is equivalent to the choice
of the linear subspace generated by $\omega\in \Gamma^{18,2}\otimes_\Z\C$.

In the duality between $M$-theory on K3 and the heterotic string on
$\S^1\times \T^2$, if we want the $\S^1$ radius to go to infinity,
we must take the area of $F$ to zero.  The remaining moduli are then
only the choice of $\omega$.  That is why, once one gets to four-dimensional
quantum field theory, with ${\cal M}$ being a piece of K3, one
sees precisely a complex structure on ${\cal M}$ in which ${\cal M}$
is elliptically fibered and no other data.

If, however, one want to get  quantum field theory on $\R^3\times \S^1$, with
a finite radius of $\S^1$, one is free  to vary the area of $F$,
while keeping fixed the volume form and complex structure.
So, as stated in (1) above, the extra modulus one gets upon compactification
on $\S^1_R$ is the ability to vary the area of the elliptic fiber in the
hyper-Kahler metric, while keeping fixed the volume form and distinguished
complex structure on the moduli space.

The relation between the radius of $\S^1_R$ and the area of the fiber $F$
can be worked out as follows.  In the duality between $M$-theory on K3
and the heterotic string on $\S^1\times \T^2$, the wrapping number of 
two-branes on $F$ is dual to the momentum along the $\S^1$.  A two-brane
wrapped on $F$ has an energy which is a multiple of the area of $F$, while
a massless particle with minimum non-zero momentum along $\S^1$ has energy
$1/R$.  So under the duality, the area of $F$ is mapped to a constant times
$1/R$, explaining the last assertion in (1) above. 

\listrefs
\end